\documentclass[]{aa}
\usepackage{graphicx}
\usepackage{psfig}
\def\xmm {{\it XMM-Newton\/}}
\def\etal {et al.\ }
\begin{document}
     \title{An \xmm\ study of the RGH~80 galaxy group}
     \author{Y.-J. Xue\inst{1}$^{,}$\inst{2}, H. B\"ohringer\inst{1} and K. Matsushita\inst{1}$^{,}$\inst{3}}
     \offprints{Y.-J. Xue, \email{yxue@mpe.mpg.de}}
     \institute{Max-Planck-Institut f\"ur extraterrestrische Physik,
                Giessenbachstra\ss e, 85748 Garching, Germany
                \and National Astronomical Observatories, 
                Chinese Academy of Sciences, Beijing 100012, China
                \and Department of Physics, Tokyo University of Science, 
                1-3 Kagurazaka, Shinjyuku-ku, Tokyo, 162-8601 Japan}
     \date{Received 28 November 2003 / Accepted 27 February 2004}
\abstract{
We present an X-ray study of the galaxy group RGH~80, observed by \xmm. 
The X-ray emission of the gas is detected out to $\sim 462h^{-1}_{50}$ kpc, 
corresponding to $\sim 0.45 r_{200}$. 
The group is relatively gas rich and luminous with respect to its temperature 
of $1.01\pm 0.01$ keV. 
Using the deprojected spectral analysis, we find that the temperature peaks 
at $\sim 1.3$ keV around $0.11r_{200}$, and then decreases inwards to 
0.83 keV at the center and outwards to $\sim 70\%$ of the peak value 
at large radii. 
Within the central $\sim 60$ kpc of the group where the gas cooling time is 
less than the Hubble time, 
two-temperature model with temperatures of 0.82 and 1.51 keV and the Galactic
absorption gives the best fit of the spectra,
with $\sim 20\%$ volume occupied by the cool component. 
We also derive the gas entropy distribution, which is consistent with the 
prediction of cooling and/or internal heating models.
Furthermore, the abundances of O, Mg, Si, S, and Fe decrease monotonically 
with radius. With the observed abundance ratio pattern, we estimate that 
$\sim 85\%$ or $\sim 72\%$ of the iron mass is contributed by SN Ia, 
depending on the adopted SN II models. 
}
\authorrunning{Xue \etal}
\titlerunning{\xmm\ study of RGH~80} 
\maketitle
\section{Introduction}
In the past decade X-ray imaging spectroscopy has become a powerful probe
of the physical conditions of the hot ($T^{>}_{\sim}10^{7}$ K) intergalactic
medium (IGM) of galaxy groups. Spatial distributions of gas temperature,
density and metallicity, as well as the structures of the gravitational
potential and dark matter halos have been measured for numerous galaxy groups.
This improves significantly our knowledge of dynamical properties and 
formation of galaxy groups. Among these, the discovery of similarity breaking 
of some IGM properties from clusters to groups is crucial for the study of 
cosmic evolution of the hot IGM in very massive halos of the universe. 
The well-known example is the deviation of the X-ray luminosity ($L_{\rm X}$) 
and temperature ($T$) relation of groups and clusters from the prediction of 
self-similar model (e.g. 
Edge \& Stewart 1991; David \etal 1993; Wu, Xue \& Fang 1999; Helsdon \& 
Ponman 2000; Xue \& Wu 2000 and references therein). The break lends
strong support to the argument that the IGM of groups was much more affected
by non-gravitational 
processes, such as radiative cooling and/or feedback of star formation, 
than the IGM of clusters. Another example is the detection of entropy
excess in clusters and groups (Ponmam, Cannon \& Navarro 1999; 
Lloyd-Davies, Ponman \& cannon 2000; Xu, Jin \& Wu 2001), which reinforces
the presence of strong non-gravitational effects especially in the inner
regions of poor clusters and groups. The reliable $L_{\rm X}-T$ relations
and the entropy profiles obtained from spatially resolved X-ray imaging 
spectroscopy may allow one for the first time to distinguish between the 
competing non-gravitational models (Voit \etal 2002, 2003; 
Mushotzky \etal 2003).

X-ray spectra of groups are dominated by emission line features. 
The wealth of characteristic emission lines provides a robust tool for 
understanding the metal enrichment processes in the IGM and 
possibly for constraining the supernova history of galaxy groups 
(e.g. Finoguenov \& Ponman 1999). However, an accurate determination 
of the metal abundances depends strongly on precise measurements of the 
temperature structure of the hot IGM in groups, as the dominated Fe L-shell 
lines are highly temperature sensitive (e.g. Buote 2002; Buote \etal 2003a). 
The high spectral resolution of \xmm\ allows us to obtain the abundances 
of O and Mg in addition to Si, S and Fe, and provides us a deep insight into 
the chemical evolution history of groups with the abundance ratio pattern 
and the metal mass to light ratios 
(e.g. Xu \etal 2002; Matsushita, Finoguenov \& B\"ohringer 2003). 

In this paper, we report the results from new \xmm\ observations of RGH~80, 
which was first identified by Ramella, Geller \& Huchra (1989) 
in the center for astrophysics redshift survey.
It was identified as an extended X-ray source in the 
{\it ROSAT} All-Sky Survey with an X-ray luminosity of 
$3.33\times 10^{43}h^{-2}_{50}$ erg s$^{-1}$ in the 0.1--2.4 keV band
within an aperture of $1.0h^{-1}_{50}$ Mpc (NRGs241; Mahdavi \etal 2000). 
Global fitting of the {\it ASCA} SIS and GIS data by 
Davis, Mulchaey \& Mushotzky (1999) 
yielded a mean gas temperature of $1.02\pm 0.05$ keV, 
an average abundance of $0.26^{+0.16}_{-0.08}Z_{\odot}$ for the $\alpha$ 
elements and $0.20^{+0.05}_{-0.06}Z_{\odot}$ for iron. 
However, Buote (2000) pointed out that a two-temperature spectral model 
provides a better fit to the {\it ASCA} spectra of RGH~80 and has a 
metallicity that is substantially higher than that obtained by the single 
temperature spectral model, $Z=0.67^{+0.42}_{-0.24}Z_{\odot}$. We will use 
the \xmm\ data to examine the spatially resolved X-ray properties of the 
group and explore their physical applications.

In section 2, we describe the observations and data reduction procedure. 
We present the detailed spectral analysis in section 3. 
In Section 4 and 5, we derive the gas density profiles, and calculate the 
three dimensional distributions of gas and dark matter, respectively. 
We discuss and summarize the implications of our results in section 6 and 7. 
Throughout this paper, we take $H_0 = 50$ km s$^{-1}$ Mpc$^{-1}$, 
$\Omega_{\rm M} =1$ and $\Omega_{\Lambda}=0$. At the group redshift of 0.037, 
$1\arcmin$ corresponds to 60.2 kpc. 
Unless stated otherwise, we adopt $68\%$ confidence limit for error analysis.
\section{Observation and data preparation}
\label{sec:prep}
RGH~80 was observed with \xmm\ for $\sim 33$ ks in revolution 563 
on January 5th, 2003. The two MOS-CCD cameras were operated 
in full frame mode, and the pn-CCD camera was operated 
in extended full frame mode. 
All cameras were covered with the Thin1 filter. 
We generated the calibrated events lists for the data by using the tasks 
{\em emchain} and {\em epchain} that are packaged in the XMM SAS v5.4.1 
software. 
In the analysis, we keep the events with PATTERNs 0--12 for the MOS cameras, 
and the events with PATTERNs 0--4 for the pn camera. 

The EPIC background is mainly composed of solar soft protons, cosmic rays, and 
cosmic X-ray photon background (Lumb \etal 2002; Read \& Ponman T.J. 2003). 
The background induced by soft protons is time variable, and hence causes large
variations of intensity (flares) in the light curves. It can be subtracted by 
discarding those flare periods. Since in the very high energy band 
the effective area of XMM is negligible and the emission is dominated 
by the particle background, we extract the light curves 
in the energy band of 10--12 keV and 10--15 keV for the MOS and pn data, 
respectively, in 100s bins. An inspection of the light 
curves does not reveal any strong temporal variations. 
In order to determine the mean count rate and its error ($\sigma$) 
for the quiescent periods, we recursively clean the light curve by 
rejecting those time intervals during which the counts are $3\sigma$ 
outside the mean value, until the mean counts remain constant. 
We then define the thresholds at the $\pm 3\sigma$ level, and reject any 
time intervals outside these thresholds. After applying this screening 
criterion, the final useful exposures are 33.0 ks for MOS1, 
32.7 ks for MOS2, and 26.6 ks for pn, respectively. 

We utilize the EPIC blank sky event files available from the XMM calibrations
 (Lumb 2002) to represent the remaining quiescent background, which is 
mainly composed of the cosmic X-ray background and the background induced 
by cosmic rays. We applied the same PATTERN selection and light curve 
screening criteria as have been used above to the blank sky events. 
Since some internal background fluorescence lines show strong spatial 
inhomogeneities, such as Al K and Si K fluorescent lines in the MOS 
cameras (Lumb \etal 2002; Read \& Ponman 2003) and 
Cu K fluorescent lines in the pn camera 
(Freyberg, Pfeffermann \& Briel 2001; Lumb \etal 2002; Read \& Ponman 2003), 
we use the {\em skycast} script 
\footnote{\textsf{http://www.sr.bham.ac.uk/xmm3/scripts.html}} to cast the 
background files into the corresponding sky coordinates of the source 
observations to ensure that the products of the background and the source 
are extracted from the same location.

Before subtracting the background, we first correct the vignetting effect 
for both source and background data sets with the SAS task {\em evigweight}, 
which computes a weight coefficient for each photon by using the inverse of 
the ratio of the effective area at the photon position and energy to the 
central effective area at the same energy. 

In this paper, we follow the method of background subtraction as proposed by 
Arnaud and collaborators (cf. Arnaud \etal 2001; 
Pratt, Arnaud \& Aghanim 2001; Arnaud \etal 2002; 
Majerowicz, Neumann \& Reiprich 2002).
Because the cosmic ray background varies from observation to observation by 
$\sim 10\%$, we normalize the background level of the blank sky files to that 
of the source observations. The ratios of the total count rates of the source 
observations to those of the blank sky data sets, extracted in the whole FOV 
of each camera in the energy band of 10--12 keV for the MOS data and 
10--15 keV for the pn data, are adopted as the background normalization 
factors. They are 1.02, 1.00 and 0.96 for MOS1, MOS2 and pn respectively, 
for this observation.  

The difference in the soft spectral component of the cosmic X-ray background 
between the source region and the blank field is investigated as follows.
We first accumulate spectra in the annular region of 10--12$\arcmin$ 
for both source and blank sky data sets. This region is centered on the 
emission centroid of the group, and is located outside the group region.
The emission centroid is computed within a $2.5\arcmin$ radius. Then we 
subtract the normalized blank sky spectrum from the source spectrum to obtain
a residual background spectrum, which is subsequently subtracted from the 
spectrum of the group after rescaling its level according to the size of 
the area where the group spectrum is extracted. 

We generate the on-axis ancillary response files (ARFs) using SAS 
task {\em arfgen}, and adopt the redistribution matrix files (RMFs) 
m11\_r7\_im\_all\_2002-11-07.rmf, 
m21\_r7\_im\_all\_2002-11-07.rmf and 
epn\_ef20\_sdY9.rmf for MOS1, MOS2 and pn respectively. 

Following the standard method 
\footnote{\textsf{see http://wave.xray.mpe.mpg.de/xmm/cookbook/EPIC\_PN/ \\ ootevents.html}}, 
we have corrected the out-of-time events of the pn data by a factor of 0.0232.

\begin{figure}
\begin{centering}
\includegraphics[scale=0.45,angle=0,keepaspectratio]{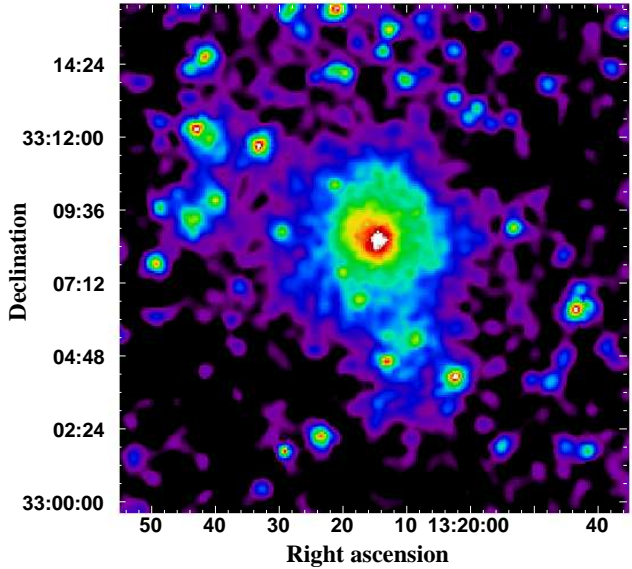}
\caption{{\footnotesize Vignetting corrected MOS image of RGH~80 
in the 0.5--3.0 keV energy band, adaptively smoothed with a signal to noise 
ratio of 10 and a maximum gaussian smoothing size of $\sigma=5$ pixels.}}
\label{fig:image}
\end{centering}
\end{figure}

We show the vignetting corrected 0.5--3.0 keV image in Figure~\ref{fig:image},
which is produced by adaptively smoothing the combined exposure-corrected MOS1
 and MOS2 image. We find that the X-ray emission of the group is extended and 
almost symmetric, except that it is contaminated by several fairly bright 
sources at 3.5--$6.5\arcmin$ from the group center. 
We then use SAS task {\em edetect\_chain} to perform source detections 
in five energy bands 
(i.e., 0.2--0.5, 0.5--2.0, 2.0--4.5, 4.5--7.5, and 7.5--12.0 keV) and 
determine the source sizes following Katayama \etal (2002), which is 
described below. 
We divide the whole FOV into 8 annular regions centered at 
the centroid of the group emission, and extract an image for each region.
The images are binned with a bin size of 40 pixels, 
and smoothed with a maximum gaussian smoothing size of $\sigma=5$ pixels.
For each annular region, the average count rate per pixel and its error 
($\sigma$) are found by recursively excluding the pixels with count rate 
$4\sigma$ above the mean value from the image until the mean count rate 
per pixel converges at a constant. Finally, for each source detected by 
{\em edetect\_chain} except for the group, we determine its size, within which
the count rate is $4\sigma$ higher than the mean value. The size of the point
sources determined by this method is $\sim 20\arcsec$. We mask out all the
sources in our further analysis for both source and blank sky data sets. 
%
%
%
\section{Spectral analysis}
\label{sec:spec}
While the diffuse X-ray emission of RGH~80 extends to a radius of 
$4.16\arcmin$ at $3\sigma$ level, 
the spectra in region $4.16\arcmin <r<7.67\arcmin$ still show 
strong features of a $\sim 0.95$ keV plasma. 
We therefore include this
region in the spectral analysis.    
We extract spectra for a series of annuli centered at the centroid of the 
group emission in such a way that for each annulus, with a width larger 
than $30\arcsec$ (PSF consideration), the MOS spectrum has at least 3000 
source counts in the 0.4--5.0 keV band to give a robust constraint 
on elemental abundances and gas temperatures. 
Finally, we obtain a total of four radially binned spectra for each camera 
(see Table 1 for details). 
The contribution from the source to the total count rate ranges from 
$\sim 97\%$ for the innermost spectrum to $\sim 20\%$ for the outermost one. 
The spectra are then grouped so that each bin has at least 20 counts, 
thereby allowing $\chi^2$ statistics to be used. 
We use XSPEC version 11.2.0 (Arnaud 1996) to analyze the spectra 
with one or two photoelectrically absorbed VAPEC spectral components 
(Smith \etal 2001) by fixing the hydrogen column density at the Galactic value 
($1.05\times 10^{20}$ atom cm$^{-2}$; Dickey \& Lockman 1990). 
For the spectra of the central region, more complicated spectral models 
such as cooling flow models are also examined. 
We adopt the solar abundances 
from Anders \& Grevesse (1989) with Fe/H$=4.68\times 10^{-5}$ 
by number, which is 1.45 times larger than the meteoritic value. 
We have taken this difference into account when comparing with other studies.
We divide the elements into six groups, 
i.e., O and Ne; Mg; Si; S and Ar; Fe and Ni; and the others 
(see also Finoguenov, Arnaud \& David 2001). 
In group 1--5, the metal abundances in the same group are tied to each other
 and are set free in the spectral fittings. In group 6, metal abundances are
fixed at the solar values. 
We perform a joint fit to the MOS1, MOS2 and pn 
spectra with the same model and model parameters, only allowing the 
normalization of each spectrum to vary independently.
%
%
\begin{table}
\begin{center}
\caption{{\small Quality of deprojected spectral fits}}
\begin{tabular}{cccccc}
\hline
\hline
       & R & \multicolumn{2}{c}{1T} & \multicolumn{2}{c}{2T}\\
 shell & (arcmin) & $\chi^2$ & dof & $\chi^2$ & dof \\
 \hline
 1 & 0.00-0.58 & 335 & 273 & 230 & 271 \\
 2 & 0.58-1.67 & 260 & 280 & 219 & 278 \\
 3 & 1.67-4.08 & 237 & 320 & 230 & 318 \\
 4 & 4.08-7.67 & 422 & 371 &     &    \\
\hline
\end{tabular}
\end{center}
\label{tab:chi}
\end{table}
%
%
\begin{figure*}[]
\begin{centering}
\includegraphics[scale=0.9,angle=0,keepaspectratio]{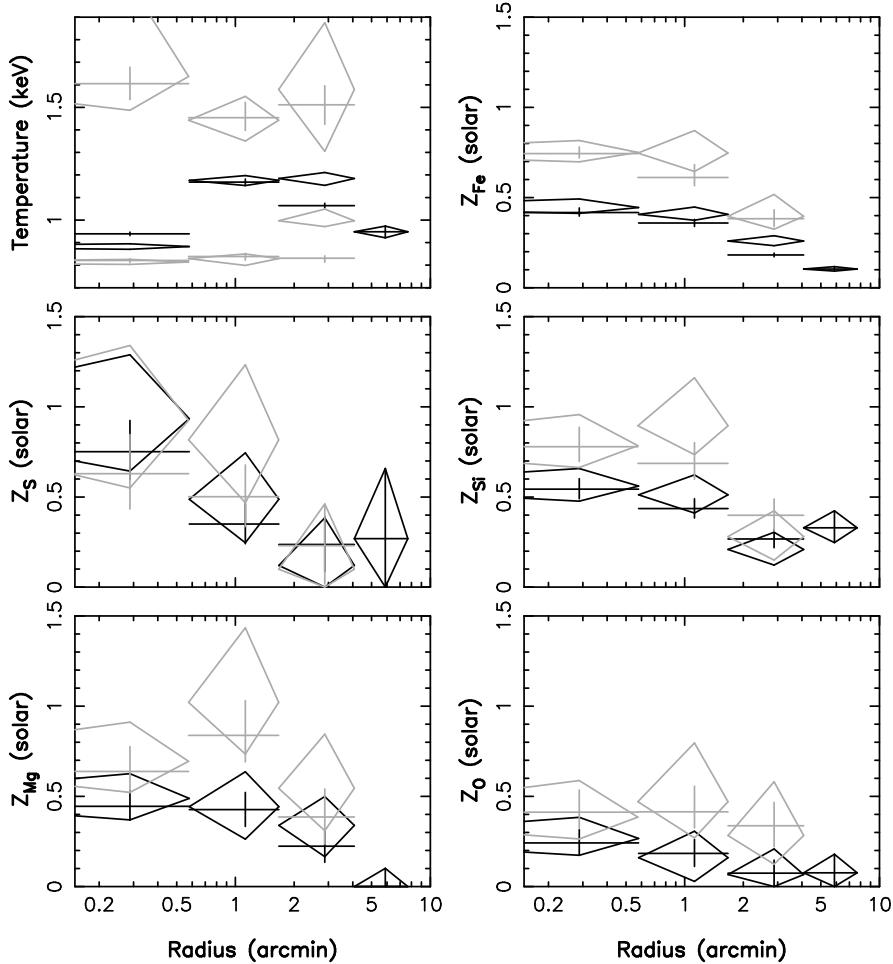}
\caption{{\footnotesize The deprojected (diamonds) and projected (crosses) temperature and abundance profiles and $1\sigma$ errors for the 1T (black) and 2T (gray) models. 
The emission measures of the hot and cool components of the 2T model are displayed in Figure~\ref{fig:2Tgas}.
}}\label{fig:vATZ}
\end{centering}
\end{figure*}
\subsection{Deprojection}
\label{sec:depjct}
With the standard ``onion peeling'' technique, we calculate the deprojected 
spectra by subtracting the emission projected from the outer shells, 
assuming that the spectral shape within each shell is the same and that the 
surface brightness profile is described by a $\beta$ model with $\beta=0.45$ 
and $r_{\rm c}=0.2\arcmin$, as will be shown in \S~\ref{sec:1Tner}. 
In order to take the contribution from beyond the outermost shell 
into account, we assume that the spectrum therein has the same shape as 
the one
in the outmost shell and has the intensity proportional to the surface 
brightness of that region (e.g. Matsushita \etal 2002).
Note that the deprojection analysis will propagate the errors of outer shells 
to inner ones. Therefore, the errors of deprojected spectra will become 
larger than those of the corresponding projected ones. The reduced 
chi-squared values of the deprojected spectral fitting become smaller than 
unity in most cases accordingly (see Table 1).
In what follows We will mainly focus 
on the analysis of the deprojected spectra to obtain the three dimensional 
X-ray properties of the group. 
\subsection{Radial temperature \& abundance profile}
\label{sec:TZ}
\subsubsection{Single temperature fits}
\label{sec:1T}
First we examine the deprojected spectra with a simple single-temperature 
VAPEC model (1T). The best-fit results are summarized in Figure~\ref{fig:vATZ}
 with black diamonds and in Table 1. We find that the temperature increases 
from $\sim 0.88$ keV at the center to $\sim 1.18$ keV at 0.6--4$\arcmin$,
and then decreases to $\sim 0.95$ keV at large radii. 
Meanwhile,
all the metal abundances decrease monotonically with radius, expect for the 
Si abundance in the outermost region that may have been biased by  
contamination of the background. 
In particular, the profile of the S abundance
shows a steeper gradient: From the center to the outer regions, 
$Z_{\rm S}/Z_{\rm Fe}$ drops from 2.10 solar to 0.46 solar. 
This relatively higher enrichment of S around the center was also found 
in M~87 with \xmm\ (Matsushita \etal 2003) and in the group NGC~1550 
with $\it Chandra$ (Sun \etal 2003).
Within $7.67\arcmin$, the emission weighted temperature is 
$1.01\pm 0.01$ keV, and the average iron abundance is $0.20\pm 0.01Z_{\odot}$,
which are in good agreement with the previous ASCA results 
($kT=1.02\pm 0.05$ keV, Fe$=0.20^{+0.05}_{-0.06}Z_{\odot}$; Davis \etal 1999). 

\begin{figure*}[]
\centerline{
\resizebox{9.cm}{!}{\psfig{figure=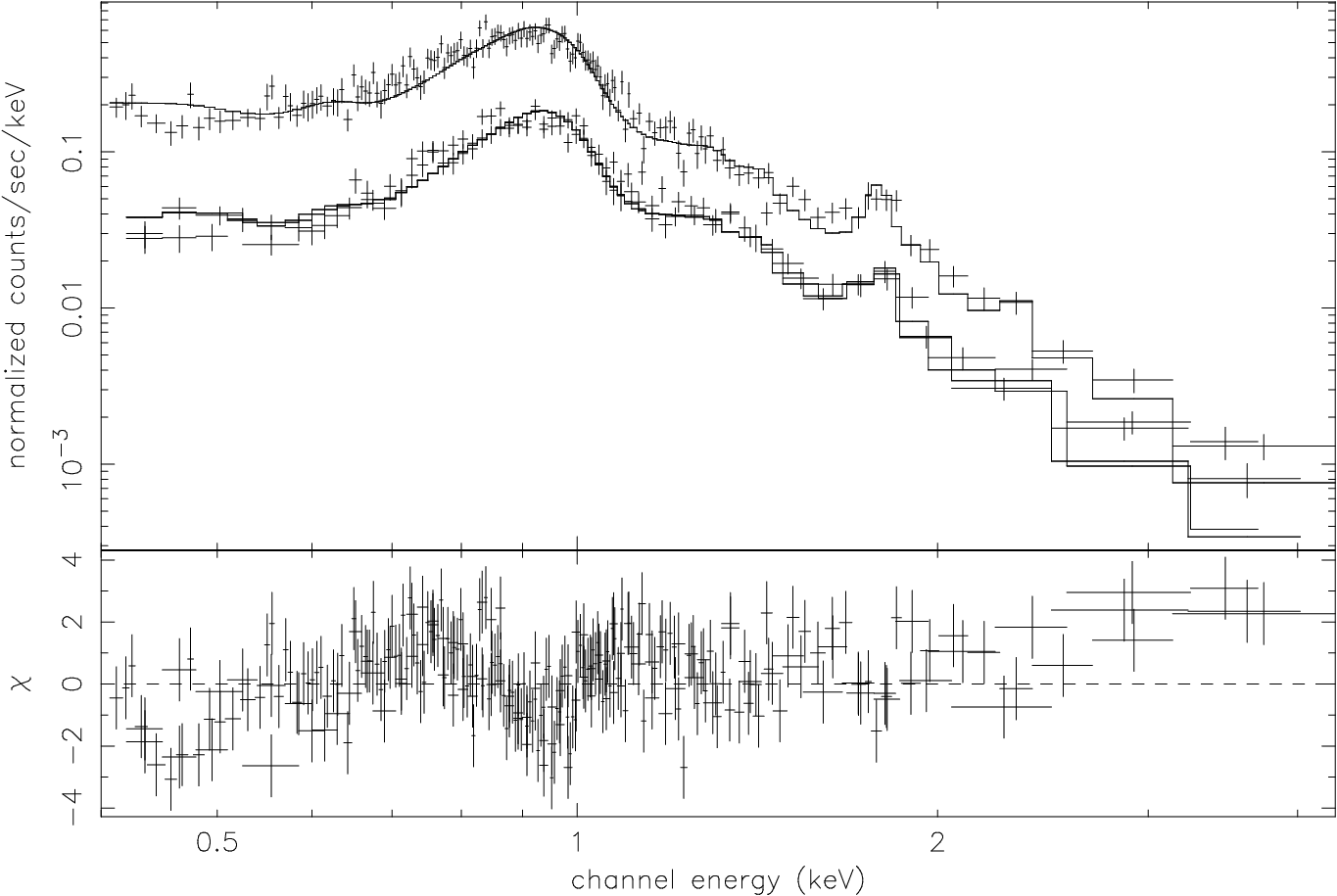,width=0.49\textwidth,angle=270,height=0.24\textheight}}
\resizebox{9.cm}{!}{\psfig{figure=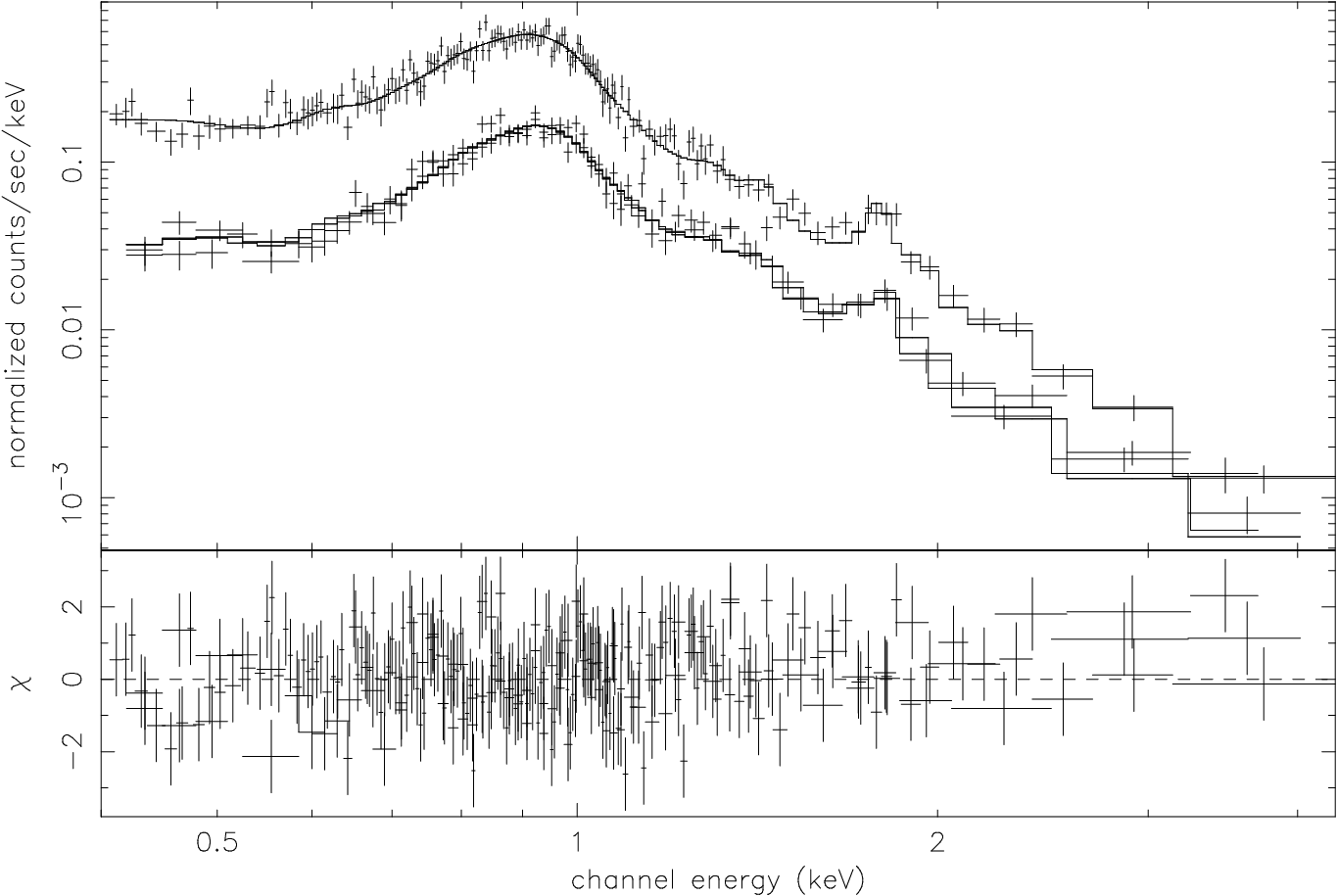,width=0.49\textwidth,angle=270,height=0.24\textheight}}
}
\caption{{\footnotesize MOS (lower crosses) and pn (upper crosses) spectra for annulus 1 fitted with the 1T (left panel) and 2T (right panel) models. }}\label{fig:spec}
\end{figure*}

It can be seen from Table 1 that the fitting of the best-fit 1T model is good 
for shells 2--4 while becomes 
much worse for the innermost region. 
As shown in Figure~\ref{fig:spec}, the poor fit is due to the large 
residuals at 0.7--1.2 keV and at above 2.5 keV where the model 
predictions are apparently below the observed data.
The former is mainly caused by the excess Fe L emissions, 
which implies the existence of cooler gas, and the latter indicates that
there should be a harder spectral component. 
This suggests the need for a multiphase IGM model (cf. Buote 2000, 2002;
Buote \etal 2003a). 

We have also made an attempt to fit the projected spectra with the 1T 
model and plotted the results in Figure~\ref{fig:vATZ} with black crosses. 
In terms of the derived gas temperature and metal abundances 
the difference between the projected and deprojected results is only minor,
with the deprojected analysis giving slightly larger abundances ($\sim 10\%$) 
for most cases.
%
%
\subsubsection{Two temperature fits}
\label{sec:2T}
In order to improve the fits to the observed spectra, we employ a multiphase 
gas model by adding another thermal spectral component to the 1T model. 
In this two temperature model (2T), the two VAPEC components are subjected 
to a common absorption that is fixed at the Galactic value. The metal 
abundances of the two components are tied to each other, while the gas 
temperature and normalization of the second thermal component are 
left free.

As shown in Table 1, the 2T model gives significantly better fits to the data
than the 1T model does, especially for the inner two shells 
(e.g. Figure~\ref{fig:spec}).
For shell 3, the 2T model improves the fitting only slightly. 
In Figure~\ref{fig:vATZ}, we display the temperature and abundance profiles 
obtained with the 2T model with gray diamonds. 
The emission measures of the hot and cool components of the 2T model are 
displayed in Figure~\ref{fig:2Tgas}. Within $2\arcmin$, the temperature of 
the cool component 
remains constant at $\sim 0.82$ keV, which is close to the 
central temperature obtained with the 1T model fit. Within 
measurement uncertainties,
the temperature of the hot component is also constant 
over the whole group region.

The best-fit of the 2T model gives significantly larger abundances than the 
1T model for O, Mg, Si and especially Fe. For example, within
the central $1.67\arcmin$, the Fe abundance obtained with the 2T model 
is $\sim 0.76Z_{\odot}$, which is nearly 2 times higher than that 
obtained with the 1T model ($\sim 0.42Z_{\odot}$). Unlike other metal
elements, the S abundance does not change notably with the new model. 
Indeed, Buote \etal (2003b) also reported a similar trend 
in group NGC~5044. They attributed the insensitivity
to the fact that the blended helium-like S triplet (2.45--2.46 keV) is
located farther away from Fe L lines ($\sim 1$ keV) than other light 
elements such as O, Mg, and Si, so that the determination of S abundance
is less affected by the change of underlying continuum induced by the new 
model.

We have also applied the 2T model to the projected spectra and shown the 
results in Figure~\ref{fig:vATZ} with gray crosses. The projected 
temperature and abundance profiles are consistent with the deprojected 
ones within $1\sigma$ errors in all shells, except that the deprojected 
temperature of the cool component of shell 3 is higher than the projected 
one by $\sim 20\%$. This may arise from the fact that the 
projected contribution of the cool component of the outer shells is removed 
from the deprojected spectrum of shell 3. 

Because in the deprojected analysis uncertainties of the abundance 
determinations are large, and because the shapes of the radial abundance 
profiles obtained with the projected and deprojected models are similar,
in what follows we will use the metal abundances obtained in the projected 
analysis to explore the abundance ratios and their constraints on the 
supernova enrichment scenarios.
It is found that the 1T and 2T models give consistent abundance ratios 
for O/Fe, Mg/Fe and Si/Fe within $1\sigma$ errors.
Moreover, these abundance ratios show no statistically significant variations 
with radius. The derived Si/Fe ratios are approximately constant 
at $\sim 1.25$ solar for the 1T model and at $\sim 1.05$ solar for 
the 2T model, respectively. The Mg/Fe ratios are around 1 solar 
for both models. On the other hand, the O/Fe ratios are subsolar. 
For example, in shell 1 the O/Fe ratios are $\sim 0.58$ and $\sim 0.55$ solar 
for the 1T and 2T models, respectively.
By contrast, the S/Fe ratios are different between the two models.
As it has already been 
shown in \S~\ref{sec:1T}, the S/Fe ratios obtained 
with the 1T model decrease significantly with radius, while the S/Fe 
ratios obtained with the projected 2T model are constant at $\sim 0.83$ solar 
within $1.67\arcmin$, and decrease slightly in the outer regions.
%
%
\begin{figure}
\begin{centering}
\includegraphics[scale=0.55,angle=0,keepaspectratio]{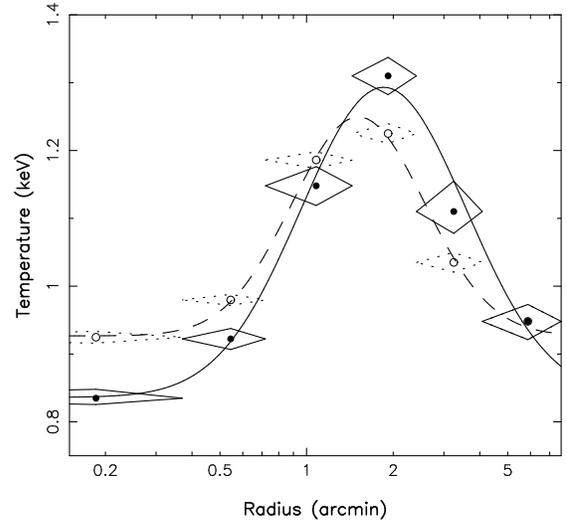}
\caption{{\footnotesize Detailed single temperature structure. Solid and dashed lines stand for the best-fit models to the deprojected (solid diamonds) and projected (dotted diamonds) temperature profiles, respectively.}}\label{fig:1TT}
\end{centering}
\end{figure}
\subsubsection{Detailed temperature structure of the IGM}
\label{sec:1TT}
In \S~\ref{sec:1T} and \S~\ref{sec:2T}, in order to obtain tight constraints 
on spatial abundance distributions, we divide the group into relatively 
few annular regions to avoid poor statistics 
in the spectral analysis, at the cost of 
losing detailed information on the temperature gradient in each shell.
In this section, we accumulate spectra for a set of narrower annuli within 
$7.67\arcmin$ (462$h^{-1}_{50}$ kpc) to perform a detailed study of the 
temperature structures. After deprojection, we fit the spectra with a single 
temperature VAPEC model, with the abundances of O, Mg, Si, S and Fe fixed 
at the best-fit values obtained in \S~\ref{sec:1T} or their linear 
interpolation. We show the resulting temperature profile in 
Figure~\ref{fig:1TT} with solid diamonds.
For comparison, we also show the results of the projected analysis with 
dotted diamonds.

We find that the temperature profile can be described by an empirical 
lognormal function:
\begin{equation}
 T(r) = T_0+\frac{A}{r/r_0}\exp{-\frac{(\ln r-\ln r_0)^2}{\omega}},
\end{equation}
in which the best-fit values are: $T_0=0.84\pm 0.01$ keV, $A=3.03\pm 0.14$ keV, $r_0=2.87\pm 0.08\arcmin$, $\omega=0.87\pm 0.04$ ($\chi_{\nu}^2$=2.0/2) and $T_0=0.93\pm 0.01$ keV, $A=1.13\pm 0.04$ keV, $r_0=2.00^{+0.04}_{-0.03}\arcmin$, $\omega=0.56\pm 0.02$ ($\chi_{\nu}^2$=0.8/2) for the deprojected and projected temperature profiles, respectively.
\subsection{Multiphase IGM in the innermost region}
\label{sec:center}
%
%
\begin{table}
\begin{center}
\caption{{\small Spectral fits within the central $1.67\arcmin$.}}
\begin{tabular}{@{}c|cccc}
\hline
Model    & A    & B$^a$    & C    & D$^b$ \\
\hline
$\chi^2$ & 440.6 & 438.0&450.9&557.9\\
$\nu$    & 441  & 441  &442  &441  \\
$T_{\rm h}$ (keV) & $1.51^{+0.07}_{-0.06}$ & $1.03^{+0.01}_{-0.01}$ & 
$1.61^{+0.04}_{-0.04}$ & $0.98^{+0.01}_{-0.01}$\\
$T_{\rm c}$ (keV) & $0.82^{+0.01}_{-0.01}$ & 0.1~(fix) & 
$0.56^{+0.02}_{-0.02}$&...\\
$EM_{\rm h}^c$& $3.27^{+0.22}_{-0.25}$ & $4.63^{+0.42}_{-0.43}$ & 
... & $5.81^{+0.67}_{-0.47}$ \\
$EM_{\rm c}^d$ or $\dot{M}^e$& $2.35^{+0.18}_{-0.09}$ & 
$39.9^{+3.9}_{-3.9}$ & $14.6^{+0.8}_{-0.8}$ &...\\
O ($Z_{\odot}$) & $0.47^{+0.14}_{-0.11}$ & $0.21^{+0.07}_{-0.06}$ & 
$0.26^{+0.09}_{-0.10}$ & $0.32^{+0.12}_{-0.05}$\\
Mg ($Z_{\odot}$) & $0.85^{+0.16}_{-0.15}$ & $0.11^{+0.07}_{-0.07}$ & 
$0.79^{+0.06}_{-0.05}$ & $0.56^{+0.11}_{-0.09}$\\
Si ($Z_{\odot}$) & $0.85^{+0.12}_{-0.09}$ & $0.30^{+0.02}_{-0.05}$ & 
$0.88^{+0.11}_{-0.12}$ &$0.58^{+0.11}_{-0.08}$\\
S  ($Z_{\odot}$) & $0.75^{+0.21}_{-0.20}$ & $0.40^{+0.13}_{-0.11}$ & 
$0.70^{+0.12}_{-0.21}$ &$0.40^{+0.20}_{-0.20}$\\
Fe ($Z_{\odot}$) & $0.76^{+0.08}_{-0.06}$ & $0.34^{+0.02}_{-0.02}$ & 
$0.70^{+0.05}_{-0.05}$ &$0.48^{+0.18}_{-0.05}$\\
\hline
\end{tabular}
\end{center}
\begin{small}
$^a$ The best-fit absorption is 
$0.31^{+0.05}_{-0.01}\times 10^{22}$ atom cm$^{-2}$.\\
$^b$ The power index and the normalization of the powerlaw component are
$1.67^{+0.14}_{-0.18}$ and $2.53^{+0.76}_{-0.88}$ photons/keV/cm$^{2}$/s, 
respectively.\\
$^c$ Emission measure of the hot component, 
expressed as $10^{-18}n_{\rm e}n_{\rm H}V/(4\pi D^2_{\rm A}(1+z)^2)$.\\
$^d$ Emission measure of the cool component of model A.\\
$^e$ Mass deposition rate of the cooling flow in unit of 
$M_{\odot} yr^{-1}$, for model B and C.
\end{small}
\label{tab:core}
\end{table}
According to our results in \S~\ref{sec:1T} and \S~\ref{sec:2T}, 
the cool component contributes up to $\sim 75\%$ and $\sim 30\%$ to 
the 0.4--5.0 keV luminosity emitted by the gas in shell 1 and shell 2,
respectively. Therefore, these regions may be important for studying the 
properties of the multiphase IGM as well as investigating the evidence 
for cooling flows. To this end, we extract spectra from the $r<1.67\arcmin$ 
region, and subtract the contributions from the outer shells with the method 
that has been described in \S~\ref{sec:depjct}. We then apply several more 
sophisticated spectral models to the data in the 0.4--7.0 keV energy band:
Model A is a two-temperature thermal model, described as WABS(VAPEC+VAPEC) 
in XSPEC. Next we replace one of the thermal components in Model A with an 
isobaric multiphase component multiplied by an absorber located at the source 
[WABS(VAPEC+ZWABS*VMCFLOW) in XSPEC; model B] to examine if the spectra 
are consistent with a multiphase cooling flow.
Model C is defined as WABS(VMCFLOW) in XSPEC. We fix the hydrogen column 
density at the Galactic value. In model A and B, the metal
abundances of the two spectral components are set to be equal. 
The higher temperature of the VMCFLOW component in model B is tied 
to the temperature of the thermal VAPEC component, while the lower one is 
fixed at 0.1 keV. In model C, the two temperatures are left free. 
As pointed out by Molendi \& Pizzolato (2001), 
the spectrum of a single-phase gas accumulated in the region where the gas
temperature gradient is large may also appear as multiphase. 
In this case, the gas is not truly multiphase, and therefore model C,
which can describe the minimum and maximum temperature of the single-phase
gas within the given region, should give a better fit to the data than model
A and B. We also examine the spectra with VAPEC+POWERLAW (model D) to check if 
the higher temperature component is associated with the central radio 
galaxies. We show the best-fits of these models in Table 2.  

It turns out that, 
model A and B fit the spectra equally well and give a slightly 
better fit ($\Delta \chi^2 \sim 10$) to the data than model C. 
However, the abundances obtained by model B are only half of the values 
given by model A. The temperature of the hot component in model A is 
nearly twice as high as that of the cool component. Previous studies of 
cooling flow clusters with $\it ASCA$ data have already found that a two 
temperature spectral model can well fit the spectra of cooling flows 
(e.g. Ikebe \etal 1999). In particular, Ikebe (2001) showed statistically 
that the ratios between the hot and cool temperatures are constant at 
$\sim 2$ for those clusters who demonstrate a very strong cool component 
in the $\it ASCA$ spectra. This was taken to imply that the central 
two-phase gas reflects the gravitational potential structure which has 
two distinct spatial scales: a main cluster component and a second
small-scale system.   
In terms of the face values alone, 
the temperatures of the two components we obtained do favor Ikebe's 
conclusion: The hot temperature ($\sim$1.5 keV) is close to the virial 
temperature of the group, and the cool one ($\sim$0.8 keV) is close to 
the kinetic temperature of stars in an elliptical galaxy.

As has been demonstrated by B\"{o}hringer \etal (2002), a cooling flow
with a broad range of temperature would result in a quite broad peak for
the blend of Fe L-shell lines. One has to either introduce an intrinsic 
absorption (model B) or leave the lower temperature cut-off to be determined 
by the spectral fitting (model C) in order to suppress the excess emission 
of the low temperature gas. In our case, model B fits the spectra better 
than model C with a large intrinsic absorption and mass deposition rate.
But the large intrinsic absorption column depths may imply that
cooling flow model is an incorrect spectral model to the data 
(Molendi \& Pizzolato 2001). 
Moreover, recent results from the RGS instrument onboard \xmm\ have provided
little evidence for the X-ray emission from the gas with temperature below 
a certain lower limiting (Peterson \etal 2001; Tamura \etal 2001; 
Kaastra \etal 2001; Xu \etal 2002; Sakelliou \etal 2002). 

If we allow the hydrogen column density to vary freely, we find that the 
fits are improved, with $\Delta \chi ^2=14$ and $22$ for model A and C, 
respectively.
Model C fits the data equally well as model A, with the temperature
varying continously from $1.49\pm 0.06$ keV to $0.56\pm 0.02$ keV.
However, the resulting hydrogen column densities of model A and C are 
unreasonably large, which are at least 6 times larger than the 
Galactic value. It is unlikely that there exits such a large 
amount of intrinsic absorber in this group. 
\section{Gas density profile}
\label{sec:ner}
In this section, we derive the radial density profile of the hot gas in 
the group, assuming that the gas is single-phase with a spatial temperature 
gradient as found in \S~\ref{sec:1TT}. For the inner region of the group, 
we will also attempt to obtain the gas distribution assuming that the IGM 
is composed of cool- and hot-phase gas.
%
%
\begin{figure}
\begin{centering}
\includegraphics[scale=0.7,angle=0,keepaspectratio]{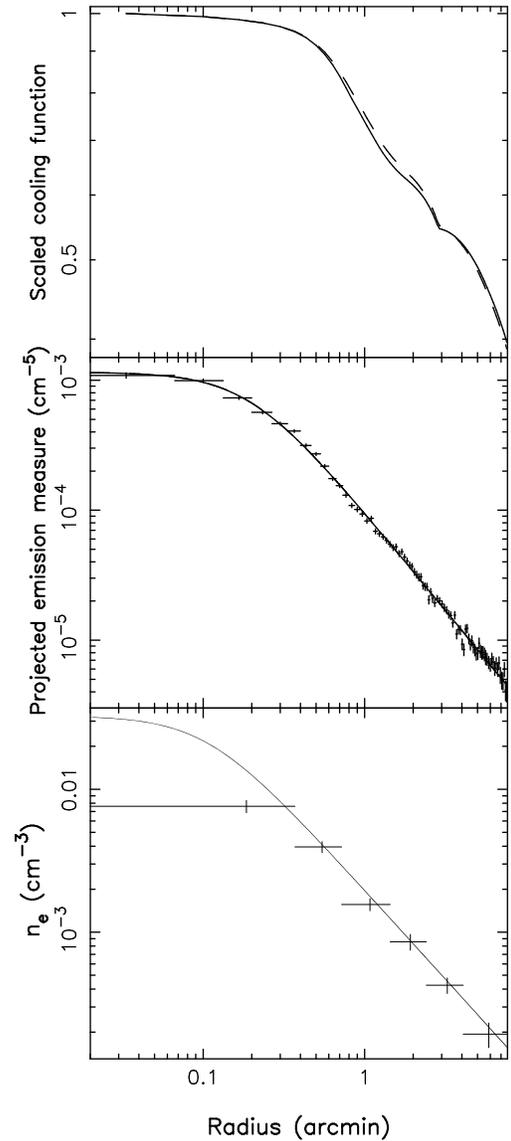}
\caption{{\footnotesize Upper panel: cooling function, $\Lambda$, profiles of MOS (dashed line) and pn (solid line), scaled to the central values. Middle panel: combined MOS and pn projected emission measure profile. See text for the detailed derivation of this profile. The solid line is the best-fit single $\beta$ model with the convolution of the \xmm\ PSF. Lower panel: electron density profiles with $1\sigma$ errors obtained from the projected emission measure (solid line) and from the normalization parameters of the deprojected 1T spectral fitting in \S~\ref{sec:1TT} (crosses).}}\label{fig:sbr}
\end{centering}
\end{figure}
%
%
\begin{figure}
\begin{centering}
\includegraphics[scale=0.7,angle=0,keepaspectratio]{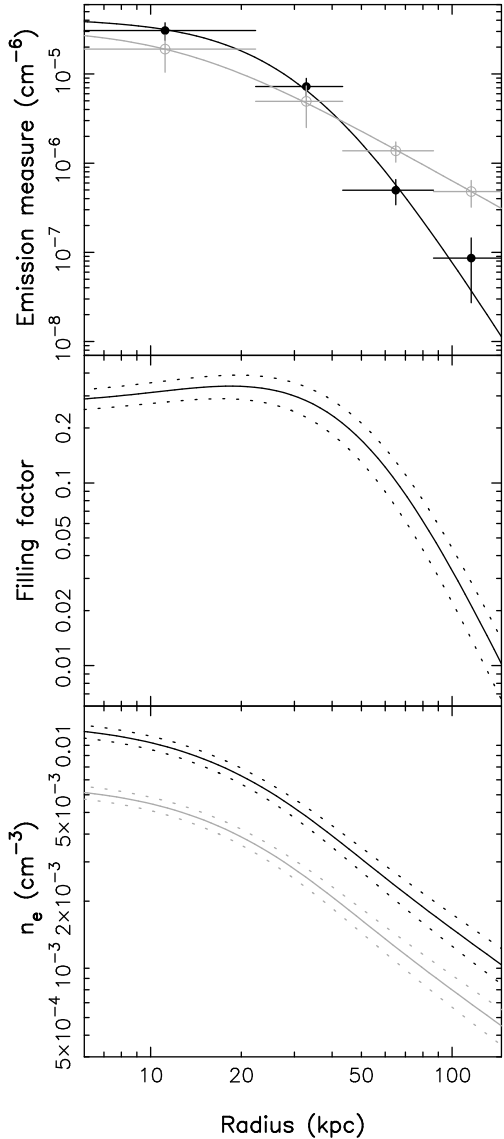}
\caption{{\footnotesize Upper panel: Emission-measure profiles of the hot (open circles) and cool (filled circles) components. Black and gray lines are the best-fit $\beta$ models. Middle panel: Filling factor profile of the cool component and the $1\sigma$ errors (dotted lines). Lower panel: electron density profiles of the hot (gray) and cool (black) components and the $1\sigma$ errors (dotted lines).}}\label{fig:2Tgas}
\end{centering}
\end{figure}
\subsection{Single-phase gas profile} 
\label{sec:1Tner}
The single-phase gas density profile can be easily obtained from the observed 
surface brightness profile. We extract the surface brightness profiles
in the 0.5--3.0 keV energy band for both source and blank sky data sets. 
The vignetting correction and background subtraction are performed using 
the same method as described in \S~\ref{sec:prep}. Since the cooling function,
$\Lambda$, depends significantly on the temperature and metal abundance
in the temperature range found in groups and cool clusters 
(see Figure~\ref{fig:sbr}), we need to take the $\Lambda$ profile into 
account when deriving the gas density profile from the observed surface 
brightness (Pratt \& Arnaud 2003). We calculate the $\Lambda$ profile by 
using an absorbed VAPEC model that is convolved with the instrumental 
response in XSPEC. The parameters in the VAPEC model are fixed at the 
best-fit projected temperature and abundance profiles, which have been given
in \S~\ref{sec:1TT} and \S~\ref{sec:1T}. From the center to the outmost 
region, the cooling function decreases with radius by $40\%$. 
Diving the observed surface brightness by the $\Lambda$ profile gives rise to 
the projected emission measure, which is proportional to 
$\int n^{2}_{\rm e}{\mathrm d}\ell$. Since the resultant projected emission 
measure profiles of MOS1, MOS2 and pn are well consistent with each other, 
we only display the averaged result in Figure~\ref{fig:sbr}. 

We fit the projected emission measure profile with the standard $\beta$ model 
(Cavaliere \& Fusco-Femiano 1976). If no correction of the XMM PSF is made,
the projected emission measure profile is well modeled by a single 
$\beta$ model with $\beta=0.419\pm 0.001$, 
$r_{\rm c}=0.195\pm 0.001\arcmin$, and 
$EM_0=1.15\pm 0.01\times 10^{-3}$ cm$^{-5}$ ($\chi_{\nu}^2=142.1/82$). 
When the convolution of the XMM PSF is performed, we find that
the fit is equally good, with $\beta=0.415\pm 0.001$, 
$r_{\rm c}=0.105\pm 0.001\arcmin$, and 
$EM_0=2.68\pm 0.01\times 10^{-3}$ cm$^{-5}$ ($\chi_{\nu}^2=142.9/82$).
Both values of $\beta$ agree nicely with the power-law relation of 
$\beta=(0.44\pm0.06)T^{0.20\pm0.03}$ found by Sanderson \etal (2003).
Note that the relatively large $\chi_{\nu}^2$ is caused by the anomalous 
fluctuations of the surface brightness profile, which can not be reduced 
even if the projected emission measure profile is fitted with a two-component 
model. 

Finally, we deproject the $\beta$ model (PSF corrected) by using the inverse 
Abell integral to calculate the gas density profile (e.g. Sarazin 1986). 
The resulting electron density profile is plotted in Figure~\ref{fig:sbr} 
with the central electron density 
$n_{\rm e0}=3.27\pm 0.01\times 10^{-2}$ cm$^{-3}$.
The $n_{\rm e0}$ is calculated from the best-fit parameters of 
the $\beta$ model using the Monte Carlo method:
We first generate $10^5$ random distributions of the parameters 
($EM_0$, $r_{\rm c}$ and $\beta$) 
around the best-fit values with the standard deviations found above, 
using a normal probability distribution. 
We then compute $n_{\rm e0}$ for each set of the random 
($EM_0$, $r_{\rm c}$ and $\beta$). 
Finally, we obtain the mean of the $10^5$ $n_{\rm e0}$ and 
its standard deviation.  
Throughout this paper, we use this method to calculate the quantities and
their errors, such as the filling factor and electron densities 
in \S~\ref{sec:2Tner}, the gas entropy, cooling time and masses 
in \S~\ref{sec:calc}, the SN type fractions and mass-to-light ratios 
in \S~\ref{sec:metal}, in which the measurement errors of all 
the related parameters are taken into account.

We have also tried to obtain the gas density profile from the normalization 
parameter, $K\propto n^2_{\rm e}V$, which is obtained in the deprojected 
spectral fittings with the 1T model in \S~\ref{sec:1TT}. The derived electron 
density is in good agreement with that obtained from the surface brightness 
profiles, except for the central region where the former is smaller than the 
latter by $\sim 43\%$. 
This again illustrates the defect of the 1T model.   
\subsection{Two-phase gas profile} 
\label{sec:2Tner}
In this section, we use the best-fit normalization parameters obtained 
in the deprojected 2T spectral fittings to calculate the gas density 
distributions for the multiphase gas in the inner part of the group. 
We reproduce four deprojected spectra within $2.4\arcmin$, and fit them
with a two temperature VAPEC model. We fix metal abundances to the abundance 
profiles found in \S~\ref{sec:2T} and assume that the temperatures of the 
hot and cool phases are spatially constant at the best-fit deprojected 
values of 1.56 keV and 0.83 keV, respectively. The emission measure of 
each phase is straightforwardly
calculated using the normalization parameters in the 
VAPEC model and is then modeled with a single $\beta$ model 
(see Figure~\ref{fig:2Tgas}),
\begin{equation}
EM(R)=EM_0\left[1+\left(\frac{R}{R_{\rm c}}\right)^2\right]^{-3\beta},
\end{equation}
where $R$ is the 3-dimensional radius. 

Similarly to 
Ikebe \etal (1999), we define the volume-filling factor of the cool 
component $f(R)$ as $EM_{\rm c}(R)=n^2_{\rm g,c}(R)f(R)$, 
where $n_{\rm g,c}$ is the gas number density of the cool component. Thus, 
the emission measure of the hot component can be expressed as 
$EM_{\rm h}(R)=n^2_{\rm g,h}(R)[1-f(R)]$, where $n_{\rm g,h}$ is the gas 
number density of the hot component. Assuming local pressure balance between 
the two phases, $n_{\rm g,c}(R)T_{\rm c}=n_{\rm g,h}(R)T_{\rm h}$, we are
able to calculate the filling factor and 
subsequent radial gas density 
distributions of the hot and cool components. As shown 
in Figure~\ref{fig:2Tgas}, $f(R)$ becomes smaller than $\sim 0.1$ 
outside $\sim 60$ kpc ($1\arcmin$), indicating that the cool gas occupies 
only a small fraction of the volume therein.
\section{Gas entropy, cooling time and mass distributions}
\label{sec:calc}
We will derive other dynamical quantities of the IGM such as 
gas entropy, cooling time, gas mass and total mass, as well as 
gas mass fraction, using the deprojected temperature profile shown 
in Figure~\ref{fig:1TT} and the radial distribution of the gas density 
given in \S~\ref{sec:1Tner} and \S~\ref{sec:2Tner}.
\begin{figure}
\begin{centering}
\includegraphics[scale=0.7,angle=0,keepaspectratio]{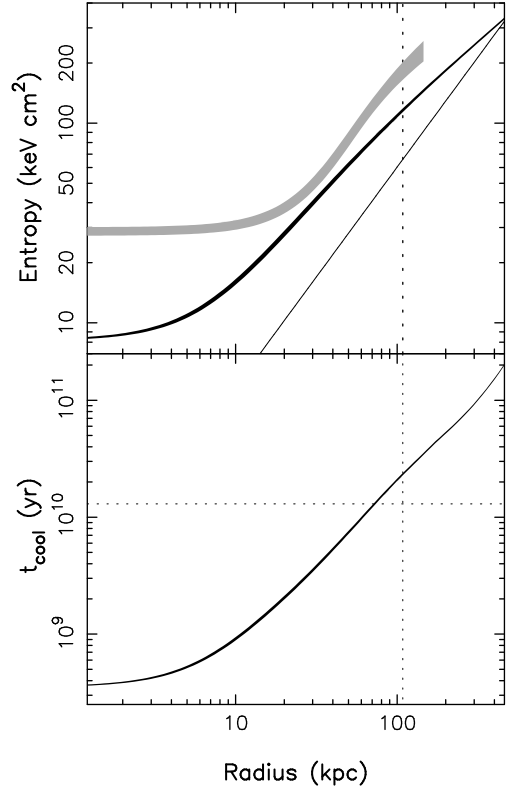}
\caption{{\footnotesize Gas entropy (upper) and cooling time (lower) profiles with $1\sigma$ errors for the single-phase gas (black region) 
and two-phase gas (gray region). The solid line represents $S(r)\propto r^{1.1}$, predicted by purely shock heating (Tozzi \& Norman 2001). The vertical dotted line marks $0.1r_{200}$, and the horizontal line marks the Hubble time.}}\label{fig:Stc}
\end{centering}
\end{figure}
\begin{figure}
\begin{centering}
\includegraphics[scale=0.7,angle=0,keepaspectratio]{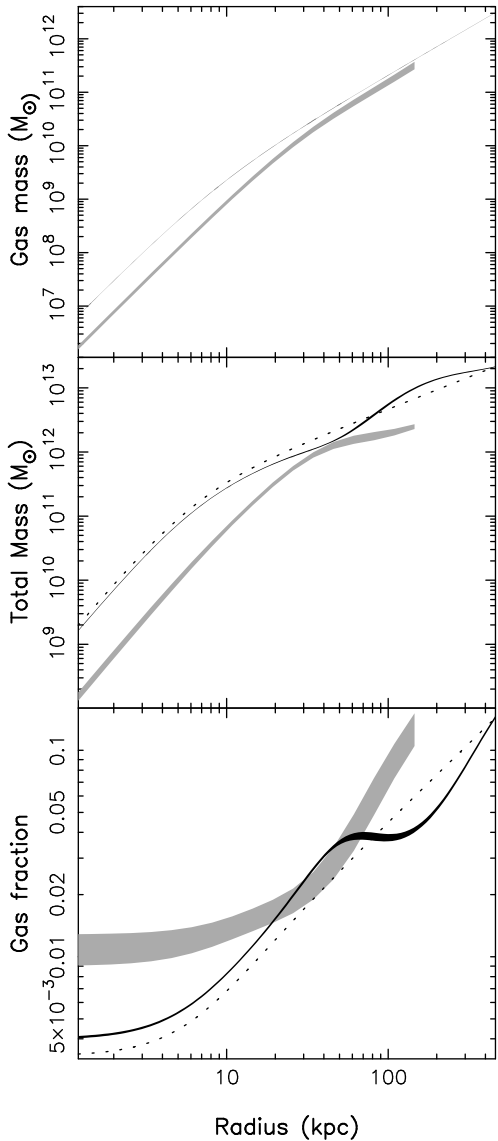}
\caption{{\footnotesize Radial profiles of gas mass, total mass and gas mass fraction of the single-phase (black) and two-phase (gray) gas with $1\sigma$ errors. Dotted lines represent the results obtained by assuming the gas is spatially isothermal.}}\label{fig:mass}
\end{centering}
\end{figure}
Following the convention in the literature (e.g. Ponman, Cannon \& Navarror 
1999), for the single-phase gas we define the gas entropy 
as $S=k_{\rm B}T/n_{\rm e}^{2/3}$, and the cooling time as 
$t_{\rm cool}=\frac{5}{2}\frac{n_{\rm g}k_{\rm B}T}{n_{\rm e}n_{\rm H}\Lambda}$. It turns out that the resulting gas entropy increases monotonically with 
radius, and does not show an isentropic floor at the central region
as far as can be resolved with the XMM instruments
(Figure~\ref{fig:Stc}). The gas cooling time 
appears to be shorter than the 
Hubble time within $\sim 80$ kpc, where the gas has been found to 
have a multiphase nature
(\S~\ref{sec:center}).
For the two-phase gas, we calculate the entropy as 
\begin{equation}
S=\exp\left[\frac{fn_{\rm e,c}\ln S_{\rm c}+(1-f)n_{\rm e,h}\ln S_{\rm h}}{fn_{\rm e,c}+(1-f)n_{\rm e,h}}\right],
\end{equation}
where the $S_{\rm c}$ and $S_{\rm h}$ are the 
gas entropies of the cool and hot components, respectively.
As seen in Figure~\ref{fig:Stc}, the derived entropy of the two-phase gas 
is larger than that of the single-phase gas by a factor of $\sim 3$ at the
center and of 1.3--2 for $R>$20 kpc.

If the IGM is single-phase, we calculate the gas mass by integrating 
the gas density over the group volume
as $\int \mu m_{\rm p}n_{\rm g}{\mathrm d}V$. 
If the IGM is two-phase, we calculate the gas mass of the cool and 
hot component using $\int \mu m_{\rm p}n_{\rm g,c}f{\mathrm d}V$ and 
$\int \mu m_{\rm p}n_{\rm g,h}(1-f){\mathrm d}V$, respectively,
where $\mu =0.6$ is the mean molecular weight,
and $m_{\rm p}$ is the photon mass.

The total gravitational mass distribtuion is derived under the assumption of 
hydrostatic equilibrium:
\begin{equation}
M(<R)=-\frac{R^2}{\rho_{\rm g}G}\frac{dP}{dR},
\end{equation}
where, $P$ is the gas pressure that is expressed 
as $P=n_{\rm g}k_{\rm B}T$ for a single-phase gas, or 
$P=n_{\rm g,c}k_{\rm B}T=n_{\rm g,h}k_{\rm B}T$ for a two-phase gas.
$\rho_{\rm g}$ is the gas mass density, defined as 
$\rho_{\rm g}=\mu m_{\rm p}n_{\rm g}$ for a single-phase gas and as 
$\rho_{\rm g}=\mu m_{\rm p}[fn_{\rm g,c}+(1-f)n_{\rm g,h}]$ 
for a two-phase gas. 

In Figure~\ref{fig:mass}, we display the spatial distributions of 
the gas mass and total gravitating mass, as well as the distribution of the
gas mass fraction. 
In 30--80 kpc, the gas mass, total mass and gas mass fraction obtained 
under the two-phase assumption are consistent with those given
by assuming that the gas is single-phase.
In the inner regions, however,
the masses derived from the two-phase assumption are 3--5 times lower
than those obtained with the single-phase assumption. 
This can be attributed to the enhanced emission of the cool component in 
the 2T model. Using the single-phase assumption, we find that the gas mass 
fraction rises rapidly with radius to
$\sim 0.15h^{-3/2}_{50}$ at $R \sim 462h^{-1}_{50}$ kpc. 

We have also
examined the effect of the temperature gradient on the 
determination of total mass. We find that the isothermal assumption 
leads to an overestimation of the total mass by $17\%$--$30\%$ within 80 kpc, 
where the gradient in temperature profile 
appears to be positive. In 80--430 kpc, where the temperature 
demonstrates a negative gradient, the isothermal assumption can 
result in an underestimation of the total mass by $48\%$ at most.
If the observed data are 
further extrapolated outside 430 kpc, the isothermal 
assumption results in an overprediction of the total mass again. 
At $R \sim 800h^{-1}_{50}$ kpc, the typical value of the virial 
radius of galaxy group at 1 keV, the overestimation is about $17\%$.

We have investigated the mass distribution models that are suggested by 
high resolution N-body simulations, such as the generalized NFW models (GNFW),
$\rho(r)=\frac{\rho_{\rm s}}{(r/r_{\rm s})^{\alpha}(1+r/r_{\rm s})^{3-\alpha}}$. The two free parameters $\rho_{\rm s}$ and $r_{\rm s}$ are determined by 
fitting the observed deprojected temperature profile with that predicted by 
assuming that the gas (described as $\beta$ model in \S~\ref{sec:1Tner}) is 
in hydrostatic equilibrium with the underlying dark matter-dominated 
gravitational potential well. Unfortunately, we fail to find an acceptable 
fit with any possible GNFW models by setting 
$\alpha =1$ (Navarro, Frenk \& White 1997, NFW), or 
$\alpha =1.5$ (Moore \etal 1999), or $\alpha =2$.
On one hand, we have only 6 data points which 
may be insufficient to give a stringent constraint on the models.
On the other hand, the temperature distribution of this group is complex, 
for example, the existence of two-phase components at the central region,
so the employment of a single-temperature assumption is over-simple.
\section{Discussion}
\label{disc}
\subsection{Gas temperature profile}
\label{sec:Tx}
Extrapolating the observed temperature and density profiles to large radii,
we can estimate the virial radius of the group, $r_{200}$, the radius 
within which the mean mass overdensity is 200 times the critical density of 
the universe.
This yields $r_{200}=786 h^{-1}_{50}$ kpc.
Such a value is, nevertheless,
$27\%$ smaller than 
the one ($r_{200}=1083 h^{-1}_{50}$ kpc) derived from the empirical relation 
$r_{200}=1.138(T/{\rm keV})^{1/2}(1+z)^{-3/2}h^{-1}_{50}$ Mpc 
found from the numerical simulations (Navarro, Frenk \& White 1995). 
Regardless of the fact that this empirical relation may significantly 
overpredict $r_{200}$ in small halos (Sanderson \etal 2003), we will
adopt $r_{200}=1083 h^{-1}_{50}$ kpc below, unless stated otherwise, 
for the convenience of comparing with other studies. In this case, 
the observed diffuse 
X-ray emission of the group extends to
$\sim 45\%$ of the virial radius.

The temperature profile of RGH~80 follows more or less a universal form as 
has already been found in some groups of galaxies observed with 
$\it ROSAT$ (Mulchaey 2000, and references therein) and 
$\it ASCA$ (Finoguenov \etal 2002a): 
Gas temperature is at its minimum ($0.83$ keV) at the center,
rises to the maximum value at a small radius ($0.11 r_{200}$), and then 
drops gradually with radius. Similar temperature profiles 
have been found recently with \xmm\ and $\it Chandra$ in galaxy groups 
NGC 1550, NGC 2563, NGC 5044, and MKW 4 (Sun \etal 2003; 
Mushotzky \etal 2003; Buote \etal 2003a; O'Sullivan \etal 2003). 
Yet, we note that there are still some groups in which a flat 
temperature profile at large radii is reported,
such as NGC 1399, NGC 2300, NGC 4325 and 
AWM 4 (Boute 2002; Mushotzky \etal 2003; O'Sullivan \& Vrtilek 2003; 
see Sun \etal 2003 for a comprehensive summary). 
\subsection{Gas metallicity and supernovae enrichment}
\label{sec:metal}
\begin{figure*}[]
\begin{centering}
\includegraphics[scale=1.,angle=0,keepaspectratio]{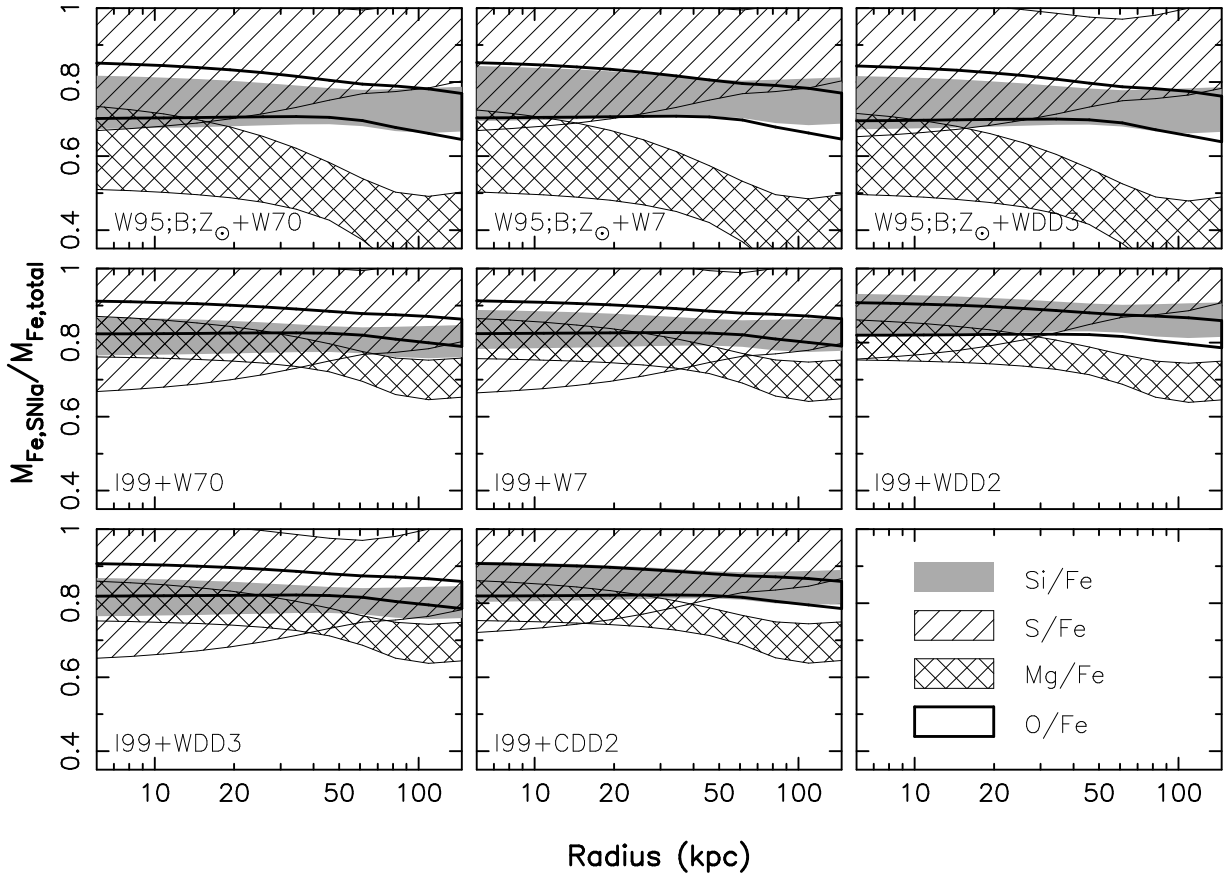}
\caption{{\footnotesize The radial profiles of SN type fractions derived from the observed abundance ratio pattern. Only the results of those models in which consistent SN type fractions can be achieved with the observed Si/Fe, S/Fe, O/Fe and Mg/Fe ratios are shown.}}\label{fig:SN}
\end{centering}
\end{figure*}
With the abundance ratios derived in \S~\ref{sec:1T} and \S~\ref{sec:2T}, 
we are able to probe the origin of the heavy elements in RGH~80 and 
to distinguish between the different contributions of type Ia and type II
supernovae. Following Gibson, Loewenstein \& Mushotzky (1997; see also 
Ishimaru \& Arimoto 1997), we take 
$M_{\rm Fe,SN Ia}/M_{\rm Fe,total}$ to quantify the roles played by 
SN Ia in contributing to the metal abundances. The relative frequency of SN Ia
 to the total SNe, $\zeta$, can be determined by the ratio of the total mass 
of the $\alpha$ elements to the total iron mass, together with the SN models. 

For SN Ia, we adopt seven updated models listed in table 3 of 
Iwamoto \etal (1999), namely, the W70 model in which the initial metallicity 
is assumed to be zero, the W7 model and five delayed detonation models, 
WDD1--3 and CDD1--2. We also utilize the SN II model listed in the same table 
(the I99 model hereafter), which is similar to the T95 model 
in Gibson \etal (1997; see also Tsujimoto \etal 1995). 
In addition, we also choose the four representative ``W95'' models 
for SN II (W95;A;$10^{-4}$Z$_{\odot}$, W95;A;Z$_{\odot}$, 
W95;B;$10^{-4}$Z$_{\odot}$, W95;B;Z$_{\odot}$; Gibson \etal 1997; 
Woosley \& Weaver 1995). 
For all the SN II models, the elemental yields are integrated 
over 10 M$_{\odot}$ to 50 M$_{\odot}$ by using the Salpeter IMF. 

We first utilize the metal abundances obtained from the spectral fittings with
the projected 2T model, in which the abundances are better constrained than
those with the 1T model at the central region. 
Based on the best-measured Si/Fe 
ratio, we find that $90\%$--$100\%$ of the iron mass was produced by SN Ia 
according to the SN Ia model WDD1/CDD1 together with any of the SN II models. 
The fraction decreases to $70\%$--$90\%$ if the other combined SN models are 
used. The SN Ia fractions estimated with the observed O/Fe ratio agree well 
with those obtained with the Si/Fe ratio. The fractions range from $65\%$ to 
$90\%$ for any combinations of the SN models.

The S/Fe ratio, on the other hand, gives a somewhat loose constraint on 
the SN Ia contribution, which ranges from $65\%$ to $100\%$ for any possible 
combinations except for those combined with WDD1 and CDD1. These two SNIa 
models (WDD1 and CDD1) seem to overproduce S relatively to Fe, which results 
in a unreasonably high SN Ia fraction ($>1$).
With the observed Mg/Fe ratio and the I99 model, we obtain a consistent 
SN Ia fraction with that derived from the O/Fe, S/Fe and Si/Fe ratios. 
However, if the four W95 models are in use, we tend to underestimate the Mg 
yields, which leads to a relatively low SN Ia fraction.  

Finally, we find that in those combined models where the SN Ia model is 
not WDD1 or CDD1, consistent SN type fractions can be achieved
with the observed abundance ratios of Si/Fe, S/Fe and O/Fe. 
If the model combination is I99+W70, I99+W7, I99+WDD2, I99+WDD3, or I99+CDD2, 
consistent SN Ia fractions can be obtained with the abundance ratios of 
Si/Fe, S/Fe, O/Fe and Mg/Fe. 
If the model combination is 
W95;B;Z$_{\odot}$+W70, W95;B;Z$_{\odot}$+W7, or
W95;B;Z$_{\odot}$+WDD3, 
a marginally consistent SN Ia fraction can be obtained in $R<20$ kpc
with the above four abundance ratios (see Figure~\ref{fig:SN}).   

%
%
\begin{table*}[]
\begin{center}
\caption{{\small The IGM metal mass to light ratios}}
\begin{tabular}{cccccccc}
\hline
\hline
Radius & $M_{\rm gas}$ & $L_{\rm B}$ & $M_{\rm Fe}/L_{\rm B}$ & 
 $M_{\rm O}/L_{\rm B}$ & $M_{\rm Mg}/L_{\rm B}$ & $M_{\rm Si}/L_{\rm B}$ &
 $M_{\rm S}/L_{\rm B}$\\
(kpc) & ($10^{12}M_{\odot}$) & $(10^{11}L_{\odot})$ & 
 $(10^{-3}M_{\odot}/L_{\odot})$ & $(10^{-2}M_{\odot}/L_{\odot})$ &
 $(10^{-3}M_{\odot}/L_{\odot})$ & $(10^{-3}M_{\odot}/L_{\odot})$ &
 $(10^{-4}M_{\odot}/L_{\odot})$\\
 \hline
462 & $3.10\pm 0.01$ & 2.89 & $3.75\pm 0.24$ & $1.19\pm 0.75$ & 
 $1.22\pm 0.45$ & $1.94\pm 0.50$ & $9.20\pm 5.33$\\
786 & $7.88\pm 0.03$ & 4.39 & $4.67\pm 0.38$ & $1.91\pm 1.56$ & 
 $1.10\pm 0.66$ & $3.00\pm 1.03$ & $13.95\pm 11.01$\\
\hline
\end{tabular}
\end{center}
\label{tab:ML}
\end{table*}

In terms of 
the radial behaviors of the SN Ia fractions derived from the 
abundance ratios 
shown in Figure~\ref{fig:SN},
it appears that the SN Ia fraction remains almost constant 
with radius at $\sim 85\%$ for the SNII model I99 or at $\sim 72\%$ 
for the SNII model W95;B;Z$_{\odot}$. 
It is likely that
the decline in the SN Ia fraction with increasing radius derived from the 
Mg/Fe ratio is an artifact because Mg lines are blended with 
the outskirt of the Fe L-shell complex, and 
therefore are measured less accurately than O and Si.
The resulting SN type fraction agrees with those obtained from 
previous $\it ASCA$ observation of galaxy groups (Finoguenov \& Ponman 1999) 
and from recent \xmm\ and $\it Chandra$ observations of NGC 5044 
(Buote \etal 2003b), NGC 1399 (Buote 2002) and NGC 1550 (Sun \etal 2003).

We then employ the abundance ratios obtained in the projected 1T model 
fittings to examine the SN models and contributions to the iron mass of
different types of SNe. We find that only in models I99+WDD1 and I99+CDD1
can a marginally consistent SN Ia fraction of $\sim 80\%$ be derived from 
the observed abundance ratios of Si/Fe, S/Fe, O/Fe and Mg/Fe.

We can calculated the ratios of the metal mass in the IGM to the total 
blue luminosity of the galaxies in the group for the measured elements
within the observed region of $462h^{-1}_{50}$ kpc 
using the elemental abundances obtained from the projected 
1T model fittings.
The metal mass of element $i$ is calculated by
\begin{equation}
M_i(<R)=\int_{0}^{R} 4\pi m_iZ_{\odot,i}Z_i(r)n_{\rm H}(r) r^2 {\mathrm d}r,
\end{equation}
where $m_i$ is the atomic mass of element $i$.
For the total blue luminosity, 
we take the magnitudes $m_{\rm zwi}$ in the Zwicky B(0) system from 
Ramella \etal 1995, and convert them to magnitude B with the relation 
$m_{\rm zwi}=B+0.35$ given by Gaztanaga \& Dalton (2000).
The metal mass-to-light ratios within the virial radius of
$786h^{-1}_{50}$ kpc are also derived by extrapolating the gas distribution
to $r_{200}$ and assuming that the abundances beyond the observed region are
the same as those in the outmost shell.
We summarize the results in Table 3. 

It turns out that the derived Fe $M/L$ is $\sim 2$
times lower than the corresponding typical 
value of clusters (Arnaud \etal 1992; Loewenstein \& Mushotzky 1996; 
Renzini 1997; Finoguenov \etal 2001). Moreover, Si $M/L$
between clusters and this group differs by a factor of $\sim$ 10.
This is consistent with the findings of Finoguenov \etal (2001).
The lower values of metal mass-to-light ratios of the group
may imply that the group has lost some of the enriched gas that was
produced by the galaxies in the group.
\subsection{Entropy}
\label{sec:S}
The observed gas entropy of RGH~80 at $0.1r_{200}$ is 
$\sim 90h^{-1/3}_{50}$ keV cm$^2$ 
for a single-phase gas model, 
which seems 
to be a typical value
for groups of $\sim 1$ keV 
(Lloyd-Davis \etal 2000; Ponman, Sanderson \& Finoguenov 2003).
The entropy reaches $\sim 185h^{-1/3}_{50}$ keV cm$^2$ if gas is 
assumed to be two-phase. 
Additionally, gas entropy increases with radius as $r^{0.81}$ 
in the region of $r>0.01 r_{\rm vir}$ and 
as $r^{1.0}$ in the region of 30 kpc$<r<$ 100 kpc
for the single-phase and two-phase assumptions, respectively. 
As a result, the derived radial entropy profiles from the single-temperature
assumption are flatter than that
expected from purely shock heating 
($S(r)\propto r^{1.1}$; Tozzi \& Norman 2001) throughout the whole group.
This implies that the IGM may have suffered from non-gravitational effects 
such as radiative cooling, heating by supernovae and/or AGNs, even at
large radii. The preheating scenario, in which gas is pre-heated and 
subsequently collapses adiabaticly, predicts an isentropic core of 
50--100 keV cm$^2$ within $0.1r_{\rm vir}$ for groups of 
$2\times 10^{13}$ M$_{\odot}$ in order to reconcile the theoretically 
expected $L_{\rm X}$--$T$ relation with observation 
(Tozzi \& Norman 2001). However, the entropy of RGH~80 we obtained starts 
to increase rapidly from an inner radius of as small as 
0.01--0.02$r_{\rm vir}$, which disagrees with the prediction of preheating 
model. 

The expected entropy profile based on radiative cooling model or 
internal heating model, or cooling plus internal heating model 
(Finoguenov \etal 2002b;
Borgani \etal 2002; Brighenti \& Mathews 2001) 
resembles the observed data of this group in shape. In particular, it has 
already been found that the cooling may marginally reproduce the observed 
scaling relations of $S(0.1r_{\rm vir})$--$T$ (Voit \& Bryan 2001) and 
$L_{\rm X}$--$T$ (Muanwong \etal 2002; Wu \& Xue 2002a), and may even be
responsible for the scale-dependence of the IGM mass function 
(Wu \& Xue 2002b). However, the cooling process suffers from the so-called
cooling crisis (Balogh \etal 2001), and is also inefficient in 
the explanation of the observed X-ray properties of groups and clusters 
(Bower \etal 2001). Inclusion of a suitable energy feedback from 
galaxy formation has been suggested to prevent 
the IGM from overcooling and supply the IGM with additional energy
(Borgani \etal 2002). Indeed, Xue \& Wu (2003) 
pointed out that a combination of cooling and heating can explain 
simultaneously the observed global X-ray properties of groups and clusters
(e.g. gas entropy distribution and $L_{\rm X}$--$T$ relation), 
and the observational limits on the contribution 
of the diffuse IGM in virialized halos to the X-ray background within 
the framework of standard CDM structure formation with an amplitude of 
matter power spectrum $\sigma_8\approx 0.7$. 
Therefore, our observation seems to favour the heating plus cooling model.
\subsection{comparison with scaling relations}
\label{sec:M}
We now compare the total mass and gas fraction of RGH~80 with 
the expectations from the scaling relations of $f_{\rm gas}$--$T$ and 
$M$--$T$ derived by Sanderson \etal (2003), adopting a virial radius 
of $786h^{-1}_{50}$ kpc.
The gas fractions at 
$0.3r_{200}$ and $r_{200}$ are $0.06h^{-3/2}_{50}$ and $0.25h^{-3/2}_{50}$, 
respectively, which are comparable to the upper limits of group gas fractions 
of in the Sanderson \etal (2003) sample. This indicates that 
RGH~80 is a relatively gas rich system. The total masses of the group 
($1.53\times 10^{13}h^{-1}_{50}M_{\odot}$ within $0.3r_{200}$ and 
$3.14\times 10^{13}h^{-1}_{50}M_{\odot}$ within $r_{200}$) agree nicely 
with what are expected from the observed $M(0.3r_{200})$--$T$ and 
$M(r_{200})$--$T$ relations of groups and clusters (Sanderson \etal 2003), 
respectively.

The velocity dispersion and bolometric X-ray luminosity of RGH~80 are
$450$ km s$^{-1}$ (Ramella \etal 2002) and 
$2.06\times 10^{43}h^{-2}_{50}$ erg s$^{-1}$ within $462h^{-1}_{50}$ kpc,
respectively. These values are comparable to the expectations from the 
observed $L_{\rm X}$--$\sigma$ relation for groups and clusters 
(Wu, Xue \& Fang 1999), though the group appears to be 
relatively luminous in the $L_{\rm X}$--$T$ plane (Xue \& Wu 2000).
\section{Conclusions}
\label{sec:conc}
We summarize below the main conclusions from our analyses of 
the \xmm\ observations of the galaxy group of RGH~80. 

\begin{itemize}
\item The X-ray emission of the group is detected out to $\sim 7.67\arcmin$, 
or $462h^{-1}_{50}$ kpc, corresponding to $0.45r_{200}$. The group seems to be 
relatively gas rich and luminous with respect to its temperature 
($1.01\pm 0.01$ keV).

\item Spectral analysis shows that the temperature profile, which increases 
from the center and then declines with radius after reaching a plateau 
around $0.11r_{200}$, follows a universal profile (Mulchaey 2000) and 
that the abundance profile of each measured element decreases 
monotonically with radius.

\item In the central region, the X-ray emission of the gas is better 
modeled by a two-temperature spectral model, with temperatures of 
0.82 and 1.51 keV and the Galactic absorption, 
than by a single temperature or cooling flow models. 
Beyond $\sim 60$ kpc ($1\arcmin$), the volume filling factor of the cool 
component, $f(R)$, becomes smaller than $\sim 0.1$, indicating that the 
cool gas occupies only a small fraction of the volume therein.

\item The gas entropy distribution derived from single-temperature
assumption deviates from the prediction of the over the whole observed region, 
The isentropic core at the center expected from preheating model does not show up. Nevertheless, our derived entropy profile resembles what is predicted by 
radiative cooling model or internal heating model or cooling plus heating 
model.
 
\item Both the abundance ratios of Fe/O and Fe/Si are very high showing
a large SN Ia dominance, which is higher than that for M87 
(Matsushita \etal 2004) and comparable to or slightly higher than that 
for Centaurus cluster (Matsushita \etal 2004).
With the abundance ratio pattern of the hot gas, we estimate that 
$\sim 85\%$ (I99) or $\sim 72\%$ (W95;B;Z$_{\odot}$) of the 
iron mass is contributed by SN Ia. This SN type fraction remains almost 
constant against radius. We find that S is significantly overproduced by 
the two delayed detonation models WDD1 and CDD1, while Mg is slightly 
underproduced by the four W95 models considered here. 

\end{itemize}

\begin{acknowledgements}
We thank B. Maughan for his continuous help in the \xmm\ data analysis,
H.-G. Xu for many constructive suggestions,
X.-P. Wu, Y.-Y Zhang and B. Qin for useful discussions, and the
referee, T. Ponman, for valuable comments. 
The present work is based on observations obtained with \xmm\, 
an ESA science mission with instruments and contributions directly funded 
by ESA Member States and the USA (NASA). 
This research has also made use of the NASA's Astrophysics Data System 
Abstract Service; the NASA/IPAC Extragalactic database (NED). 
This work was supported by MPG-CAS exchange program and by 
the National Science Foundation of China, under Grant No. 10233040.

\end{acknowledgements}

\end{document}